\newcommand{\beq}{\begin{equation}}
\newcommand{\eeq}{\end{equation}} 
\newcommand{\bea}{\begin{eqnarray}}
\newcommand{\eea}{\end{eqnarray}}

\newcommand*{\Dsl}[0]{{\rlap{\kern2.25pt /}{D}}}

\renewcommand{\d}{\delta}

\newcommand{\tK}{\widetilde{K}}

\renewcommand{\a}{\alpha}

\newcommand{\tr}{\text{Tr}}

\newcommand{\bx}{\mathbf{x}}

\newcommand{\vx}{{\vec{x}}}
\newcommand{\vy}{{\vec{y}}}

\newcommand{\vk}{{\vec{k}}}

\newcommand{\e}{\epsilon}

\newcommand{\D}{\Delta}

\newcommand{\ra}{\rightarrow}

\newcommand{\oh}{\frac{1}{2}}

\newcommand{\dg}{\dagger}
\newcommand{\non}{\nonumber}

\documentclass{PoS}
\usepackage{amsmath,amssymb,amsthm,bm,bbm,color}
\usepackage{graphics,subfigure,rotating}

\title{Polyakov line actions from SU(3) lattice gauge theory with dynamical fermions: first results via relative weights}

\ShortTitle{Polyakov line actions from SU(3) lattice gauge theory with dynamical fermions}

\author{\speaker{Roman H\"ollwieser}$^{1,2}$ and Jeff Greensite$^3$\\
E-mail: \email{hroman@kph.tuwien.ac.at}, \email{greensit@sfsu.edu}\\
$^1$Department of Physics, New Mexico State University, PO Box 30001,
Las Cruces, NM 88003-8001, USA\\
$^2$Institute of Atomic and Subatomic Physics, Vienna University of
Technology, Operngasse 9, 1040 Vienna, Austria\\
$^3$Physics and Astronomy Department, San Francisco State University, San
Francisco, CA~94132, USA}

\abstract{We apply the relative weights method to extract an effective
	Polyakov line action, at finite chemical potential, from an underlying SU(3)
	lattice gauge theory with dynamical fermions. The center-symmetry breaking
	terms in the effective theory are fit to a form suggested by the
	hopping-parameter expansion, and the effective action is solved at finite
	chemical potential by a mean field approach.  We present preliminary results
	for one-link staggered fermions with mass $ma=1.0$ and Wilson gauge action 
	at $\beta=5.4$ on $L^3\times4$ lattices with $L=16$.
\thanks{The numerical simulations were performed at the Phoenix and Vienna
Scientific Cluster (VSC) at VUT and the Riddler Cluster at NMSU. This research
was supported by the Erwin Schr\"odinger Fellowship program of the Austrian
Science Fund FWF (``Fonds zur F\"orderung der wissenschaftlichen Forschung'')
under Contract No. J3425-N27 (R.H.) and the U.S.\ Department of Energy under
Grant No.\ DE-FG03-92ER40711 (J.G.).}}

\FullConference{The 33rd International Symposium on Lattice Field Theory\\
                 14 -18 July  2015\\
                 Kobe International Conference Center, Kobe, Japan}

\begin{document}

\section{Introduction}
We present our approach to the sign problem in QCD, see {\it
e.g.}~\cite{deForcrand:2010ys,Aarts:2013bla}, which is to map QCD with a
chemical potential into a simpler effective theory, namely, the effective
Polyakov line action (PLA). The effective Polyakov line action $S_P$ can be
computed analytically from the underlying lattice gauge theory at strong gauge
couplings and heavy quark masses, and at leading order it has the form of an
SU(3) spin model in $D=3$ dimensions ($S_P$ has been computed to higher orders
in the combined strong-coupling/hopping parameter expansion
in~\cite{Fromm:2011qi}). This model has been solved at finite chemical potential
$\mu$ by several different methods, including the flux
representation~\cite{Mercado:2012ue}, stochastic
quantization~\cite{Aarts:2011zn}, reweighting~\cite{Fromm:2011qi}, and the mean
field approach~\cite{Greensite:2012xv}. The motivation is that since the phase
diagram for $S_{spin}$ has been determined over a large range of $J,h,\mu$ by
the methods mentioned above, perhaps the same methods can be successfully
applied to solve $S_P$, providing that theory is known in the parameter range
(of temperature, quark mass, and chemical potential) of interest.  The phase
diagram of the effective theory will mirror the phase diagram of the underlying
gauge theory.

Here we compute $S_P$ from an underlying $SU(3)$ lattice gauge theory at $\beta=5.4$ with staggered fermions of mass $ma=1.0$ on $L^3\times4$ lattices with $L=16$, using the ``relative weights'' method introduced in~\cite{Greensite:2013yd,Greensite:2013bya}. We then deal with the sign problem via mean field theory, as shown in~\cite{Greensite:2012xv}, which now seems to be a surprisingly accurate method for solving effective actions of this kind.

\section{The relative weights method}

The effective Polyakov line action $S_P$ of a lattice gauge theory is defined by
integrating out all degrees of freedom of the lattice gauge theory, under the
constraint that the Polyakov line holonomies are held fixed.  It is convenient
to implement this constraint in temporal gauge, where all  timelike link variables are set equal to the identity except on some time slice, {\it i.e.} ${U_0(\bx,t\ne 0)=\mathbbm{1}}$. Then the timelike link variables at time slice $t=0$ are the Polyakov line holonomies, and we have
\bea
\exp\Bigl[S_P[U_{\vx}]\Bigl] =    \int  DU_0(\vx,0) DU_k  D\phi ~ \left\{\prod_{\vx} \d[U_{\vx}-U_0(\vx,0)]  \right\}
 e^{S_L} \ ,
\label{S_P}
\eea
where $\phi$ denotes any matter fields, scalar or fermionic, coupled to the gauge field, and $S_L$ is the lattice action (note that we adopt a sign convention for the Euclidean action such that the Boltzmann weight is proportional to $\exp[+S]$).  

It is difficult to carry out the integration on the rhs of Eq.(\ref{S_P}) directly.  The relative weights approach is based on the observation that if $U'_\vx$ and $U''_\vx$ are two Polyakov line configurations which are nearby in configuration space, then the ratio
\bea
\exp[\D S_P] = {\int  DU_k  D\phi ~  e^{S'_L} \over \int  DU_k  D\phi ~  e^{S''_L} }
={\int  DU_k  D\phi ~  \exp[S'_L-S''_L] e^{S''_L} \over \int  DU_k  D\phi ~  e^{S''_L} }
= \Bigl\langle  \exp[S'_L-S''_L] \Bigr\rangle
\label{eq:rw}
\eea
is easily computed as an expectation value, by standard methods.  Here
$S'_L,S''_L$ correspond to the lattice action with timelike link variables
$U_0(\vx,0)$ at $t=0$ fixed to $U'_\vx,U''_\vx$ respectively.  Let
$U_\vx(\lambda)$ be a path through Polyakov line configuration space,
parametrized by $\lambda$. Defining	$U'_\vx=U_\vx(\lambda_0+\Delta\lambda/2)$ and
$U''_\vx=U_\vx(\lambda_0-\Delta\lambda/2)$ we can easily compute $\D S_P =
S_P[U'_\vx]-S_P[U''_\vx]$ and we are able to extract the directional derivative
$dS_P/d\lambda$ at any point along the path.  By a judicious choice of
derivatives, it is possible to compute $S_P$ itself at zero (or at finite
imaginary) chemical potential.  

\section{The effective Polyakov line action}

For heavy quarks one can use the hopping parameter expansion to expand the log
of the fermionic determinant in terms of loops. The terms which depend on chemical
potential are those terms which wind around the lattice, and at each order in
fugacity $\e^{\mu/T}$, to leading order in the hopping parameter $\kappa$,
these are the traces of Wilson lines winding multiple times around the lattice
in the time direction. If we consider only the single winding terms (which are
simply Polyakov lines), we have the contribution
\bea
h\sum_\vx(e^{\mu/T} \tr U_\vx + e^{-\mu/T}\tr U^\dg_\vx)
\label{convert}
\eea
where $U_\vx$ is the Polyakov line holonomy ($P_\vx = \frac{1}{3}\tr U_\vx$),
with $h = (2\kappa)^N_t$, with $\kappa$ the hopping parameter for Wilson fermions,
or $1/2m$ for staggered fermions, and $N_t$ is the extension of the lattice in the
time direction. In~\cite{Fromm:2011qi} it is shown that one can go a little further and sum up all the multiple winding terms which gives us, to leading order in $\kappa$, the piece of the fermionic contribution that depends on $\mu$:
\bea
\exp[S_F(\mu)] = \sum_\vx \det[1+he^{\mu/T}\tr U_\vx]^p\det[1+he^{-\mu/T}\tr U^\dg_\vx]^p
\label{eq:hop}
\eea
where determinants can be expressed entirely in terms of Polyakov line operators, using the
identities
\begin{align}
\det[1+he^{\mu/T}\tr U_\vx]&=1+he^{\mu/T}\tr U_\vx+h^2e^{2\mu/T}\tr
	U^\dg_\vx+h^3e^{3\mu/T},\non\\
\det[1+he^{-\mu/T}\tr U^\dg_\vx]&=1+he^{-\mu/T}\tr U^\dg_\vx+h^2e^{-2\mu/T}\tr U_\vx+h^3e^{-3\mu/T}.
\label{eq:hop}
\end{align}
This is the full fermionic contribution for heavy ($\kappa \ll 1$) and dense $e^{\mu/T} \gg 1$ quarks.
The power is $p = 1$ for four flavors of staggered fermions, and $p = 2N_f$ for
$N_f$ flavors of Wilson fermions. 
The parameter $h$ depends on the bare quark mass and the gauge coupling. We are
going to make the strong assumption that this form of the fermionic determinant remains approximately valid up to
the high-$T$ transition or crossover, for some choice of $h$, even in a regime  where the
hopping-parameter/strong-coupling expansion breaks down. Another approximation
is that we are neglecting center symmetry-breaking products of Polyakov lines
$P_\vx P_\vy$ (and complex conjugates) at $\vx\ne\vy$, which are generated by double-winding, 
non-overlapping Polyakov loops. 
Taken everything together, the full effective action motivated by the heavy-dense
action, that we want to determine from relative weights, should look like this:
\bea
S_{eff}[U_\vx] &=& \sum_{\vx,\vy} P_\vx K(x-y) P_\vy\non\\
&+&p\sum_\vx\log(1+he^{\mu/T}\tr[U_\vx]+h^2e^{2\mu/T}\tr[U_\vx^\dagger]+h^3e^{3\mu/T})\non\\
&+&\log(1+he^{-\mu/T}\tr[U_\vx]+h^2e^{-2\mu/T}\tr[U_\vx^\dagger]+h^3e^{-3\mu/T})
\label{eq:SP}
\eea

We first run a standard Monte Carlo simulation, generate a configuration of
Polyakov line holonomies  $U_\vx$, and compute the Polyakov lines $P_\vx$. Then,
we look at the Fourier (or ``momentum'') components $a_\vk = a^R_\vk + i
a^I_\vk$ of Polyakov line configurations $P_\vx = \sum_\vk  a_\vk e^{i \vk \cdot \vx}$,
and compute the derivatives with respect to the real part of $a_\vk$
\beq
           {1\over L^3}\left( {\partial S_P \over \partial a^R_{\vk}}\right)_{a^R_\vk = \a} \ ,
\label{eq:O}
\eeq
as a function of the lattice momentum
           $k_L = 2 \sqrt{ \sum_{i=1}^3 \sin^2(k_i/2)}$ ,
with components $k_i=2\pi m_i/L$, using the following triplets $\vec{m}=(m_1 m_2 m_3)$ of mode numbers:
\bea (000),(100),(110),(200),(210),(300),(311),(400),(322),(430),\non\\(333),(433),(443),(444),(554).\label{eq:ks}\eea

\noindent We set the momentum mode $a_\vk=0$ in this configuration to zero, to obtain the configuration 
\beq
             \widetilde{P}_\vx =  P_\vx - \left({1\over L^3} \sum_\vy P_\vy e^{-i\vk \cdot \vy}\right) e^{i\vk \cdot \vx} \ .
\eeq
Then we define
\bea
            P''_\vx = \Bigl(\a - \oh \D \a \Bigr) e^{i\vk \cdot \vx} + f
\widetilde{P}_x \qquad\mbox{and}\qquad
            P'_\vx = \Bigl(\a + \oh \D \a \Bigr) e^{i\vk \cdot \vx} + f \widetilde{P}_x \ ,
\eea
where $f$ is a constant close to one; in practice we have used $f=0.8$, the
choice $f=1$ is only possible in the large volume, $\a \ra 0$ limit.  From
$P''_\vx, P'_\vx$ we construct the holonomy configurations $U''_\vx, U'_\vx$,
and compute the
observable (\ref{eq:O}) via the relative weights method specified in Eq.
(\ref{eq:rw}), keeping the holonomies fixed during Monte Carlo simulations. We
average results over 2000 measurements on 20 different configurations for each
$k$ value in (\ref{eq:ks}) and different $\alpha=0.01-0.06$.

\begin{figure}[h]
	\centering
	a)\includegraphics[width=.47\linewidth]{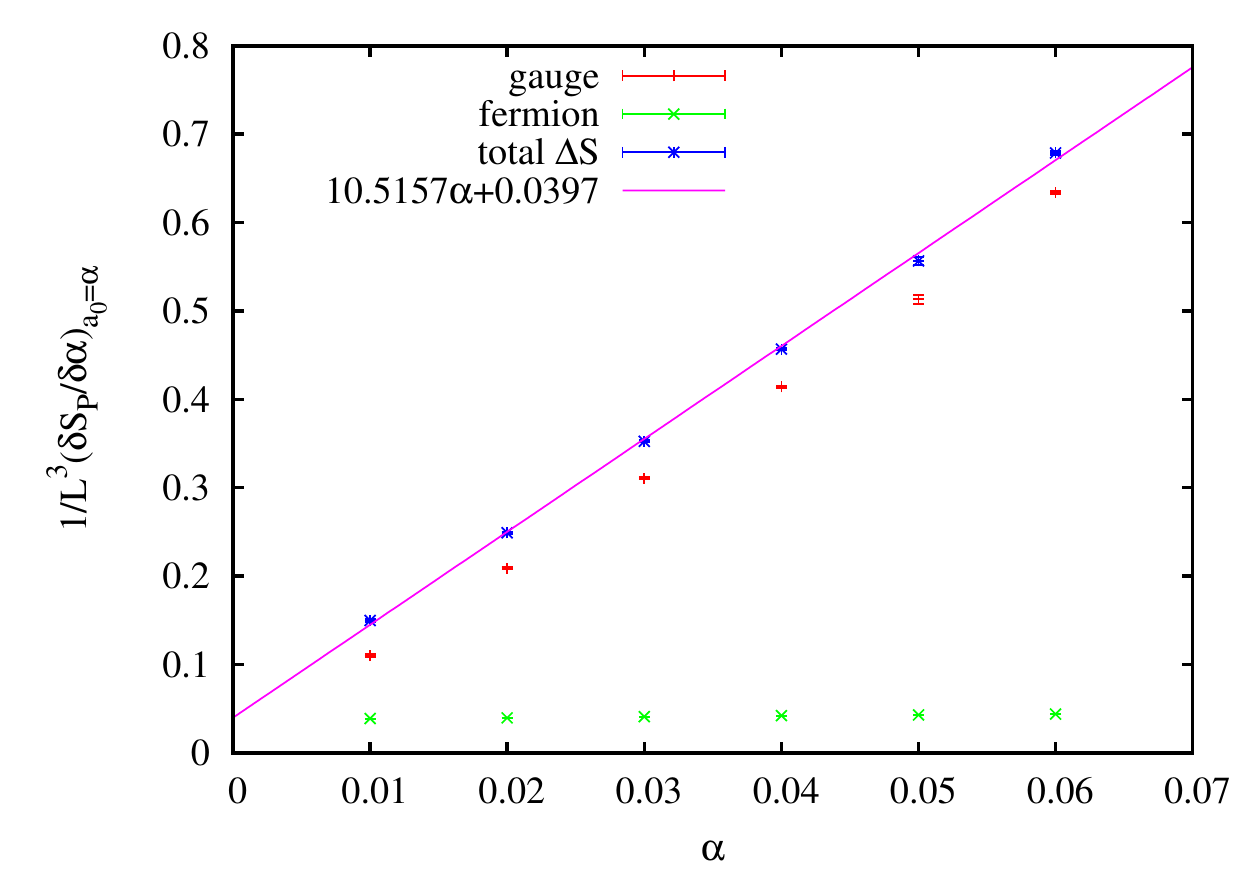}
	b)\includegraphics[width=.47\linewidth]{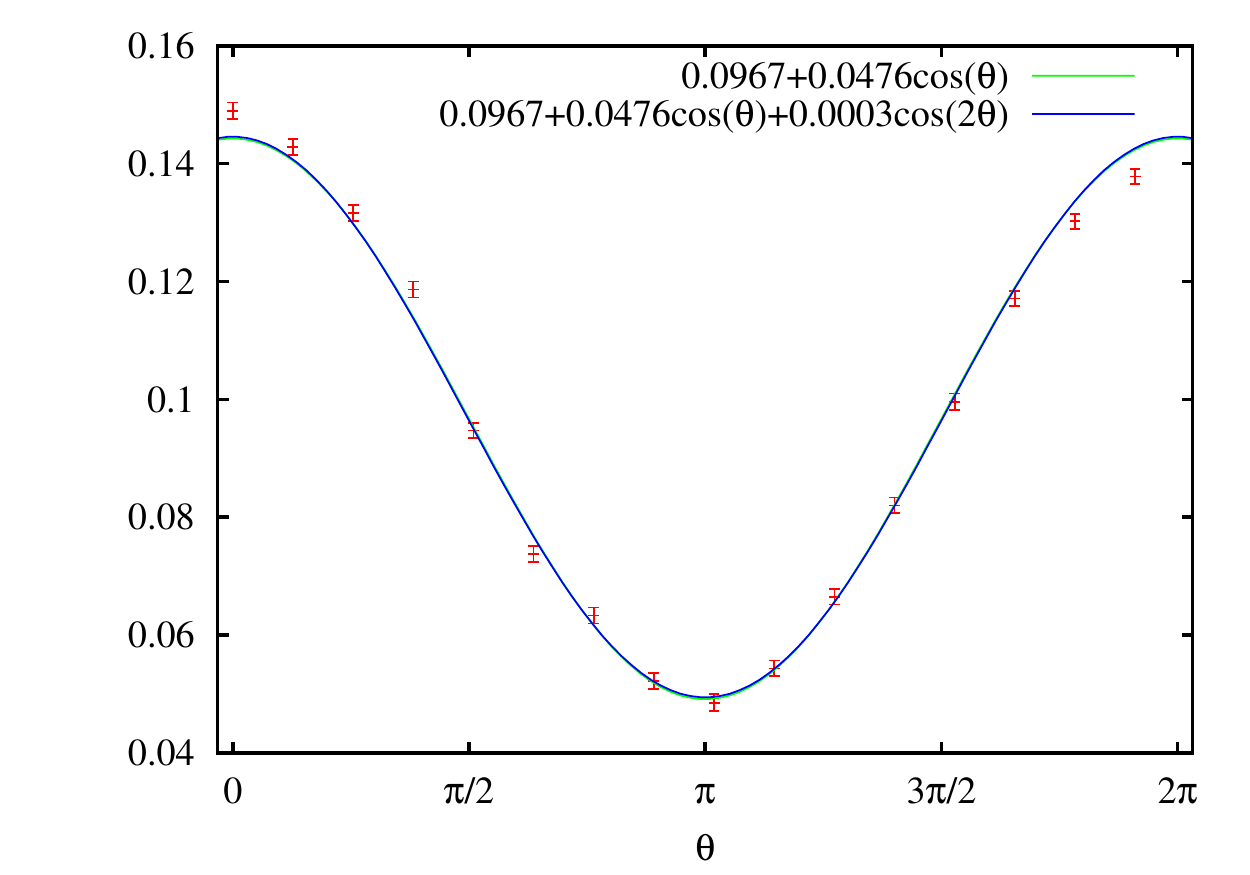}
\caption{Derivatives of $S_P$ with respect to momentum modes $a_\vk$,
	evaluated at $a_\vk=\a$ for the zero mode $k=0$: (a) data points fit by a
	straight line of slope $2\tK(0)\alpha$, (b) plotted against the imaginary
	chemical potential $\mu/T=i\theta$; note that the negligible $\cos(2\theta)$ term
	supports the neglect of center symmetry-breaking terms, at $ma=1$, which are quadratic 
	in the Polyakov lines.}
	\label{fig:dSk0}
\end{figure}

\begin{figure}[h]
	\centering
	a)\includegraphics[width=.47\linewidth]{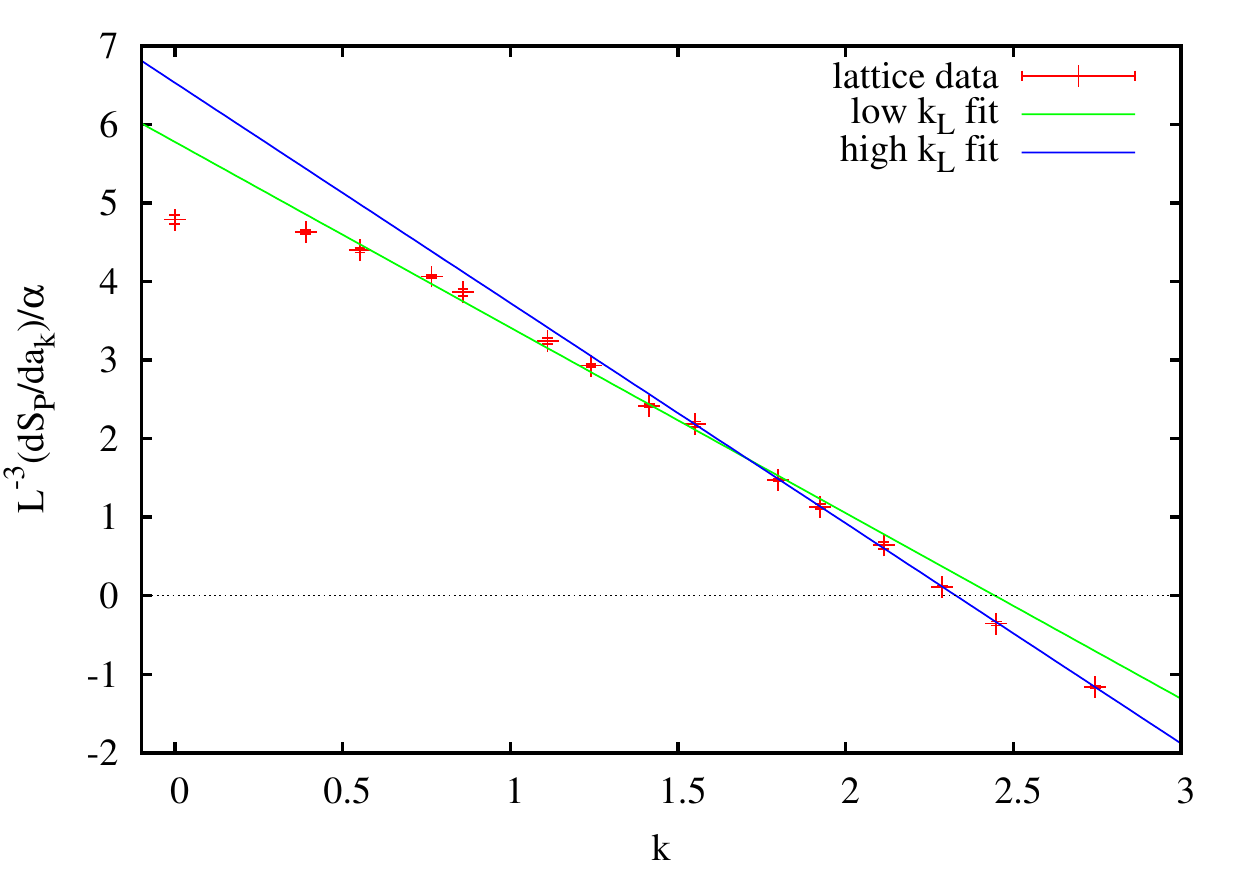}
	b)\includegraphics[width=.47\linewidth]{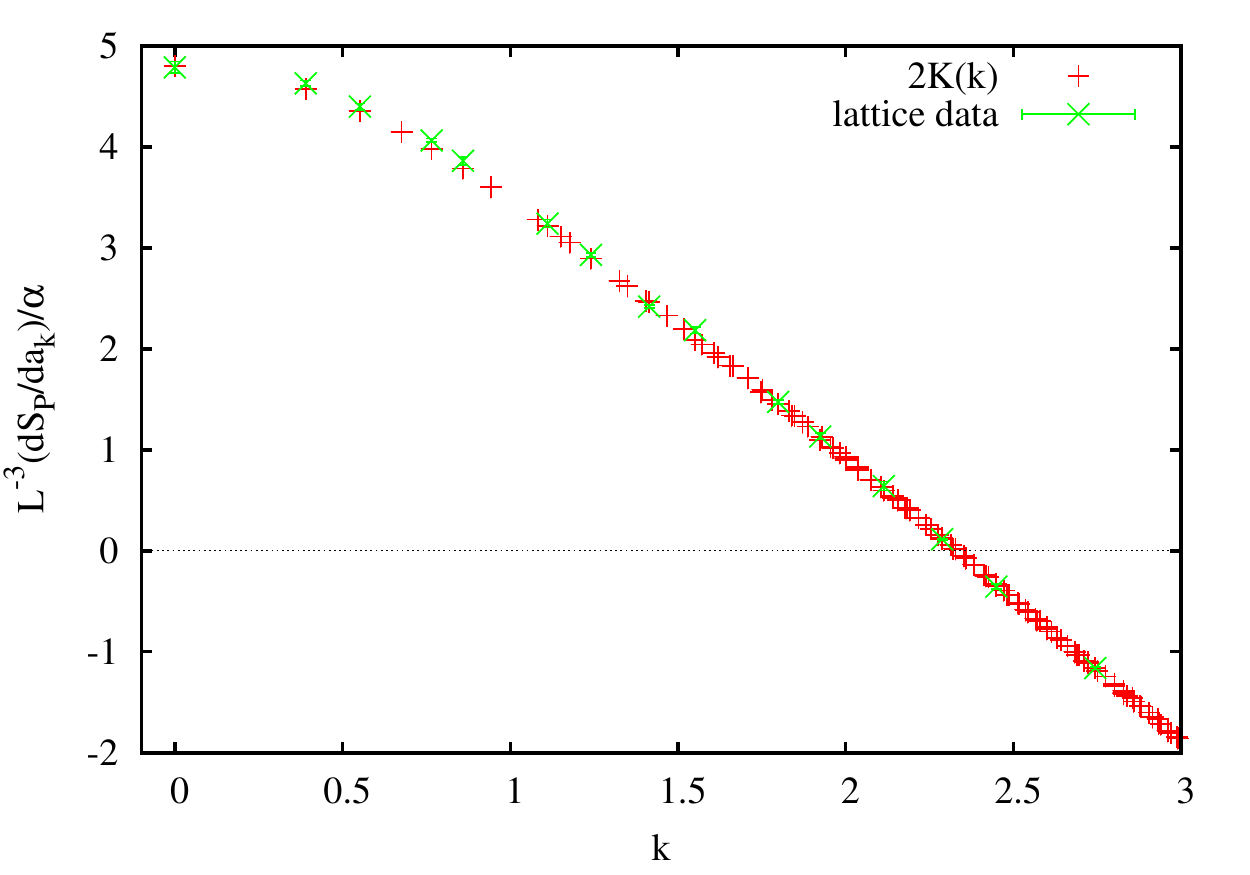}
\caption{Derivatives of $S_P$ with respect to momentum modes $a_\vk$,
	evaluated at $a_\vk=\a$ for 15 values of $k_L$: (a) data points fit by two
straight lines. (b) the data points together with $2\tK(k_L)$, determined by the
procedure explained in the text.}
	\label{fig:ks}
\end{figure}

In order to extract the kernel $K(x-y)$ we start with fitting the kernel
$\tK(k)$ in momentum space and $h$ to the lattice data for the $k=0$ mode, 
as shown in Fig.~\ref{fig:dSk0}a) via
\bea
\frac{1}{L^3}(\frac{\partial S_P}{\partial
a_\vk})_{a_\vk=\alpha}=2 \tK(k)\alpha+\frac{p}{L^3}\sum_x(3he^{ikx}+3h^2e^{-ikx} +
c.c.).\label{eq:fit}
\eea
For a more precise extraction of $\tK(0)$ and $h$ we also carry out the relative
weights calculation in a lattice gauge theory with an imaginary chemical
potential $\mu/T=i\theta$. This is done by simply multiplying the fixed
configurations $U'_\vx,U''_\vx$ of timelike links at $t=0$ by an
$\vx$-independent phase factor $e^{i\theta}$, and calculating the derivatives
of $S_P$ wrt $\theta$ at a set of $\theta$ values and fixed $\alpha$ values (see
Fig.~\ref{fig:dSk0}(b) for $\alpha=0.01$). We fit this data for each $\alpha$ to
\bea
2\tK(0)\alpha+\frac{p(3he^{i\theta}+3h^2e^{2i\theta})}{L^3\sum_x(1+3he^{i\theta}P_\vx+3h^2e^{2i\theta}P_\vx^\dg+h^3e^{3i\theta}} + c.c.
\eea
and extrapolate the results for $h$ and $\tK(0)$ to $\alpha\ra0$. Having computed
$h$ we use Eq. (\ref{eq:fit}) again to fit $\tK(k)$ for $k\ne0$ to the lattice
data at zero chemical potential again. We find that the extracted data of the
kernel $\tK(k)$ almost fits two straight lines, defined by some $\tK^{fit}(k_L)$,
except for $\tK(0)$, as shown in Fig.~\ref{fig:ks}a). 
Next, we calculate the position-space kernel with a long distance cutoff $r_{max}$
\beq
    K(\vx-\vy) = \left\{ \begin{array}{cl}
                   {1\over L^3}\sum_\vk \tK^{fit}(k_L) e^{i\vk\cdot (\vx-\vy)} & |\vx-\vy| \le r_{max} \cr \\
                      0 & |\vx-\vy| > r_{max} \end{array} \right. \ .
\eeq
The cutoff $r_{max}$ is chosen so that, upon transforming {\it this} kernel back
to momentum space, the resulting $\tK(k)$ also fits the low-momentum data at
$k=0$, as shown in Fig.~\ref{fig:ks}b).

\begin{figure}
\centering
\includegraphics[width=.49\linewidth]{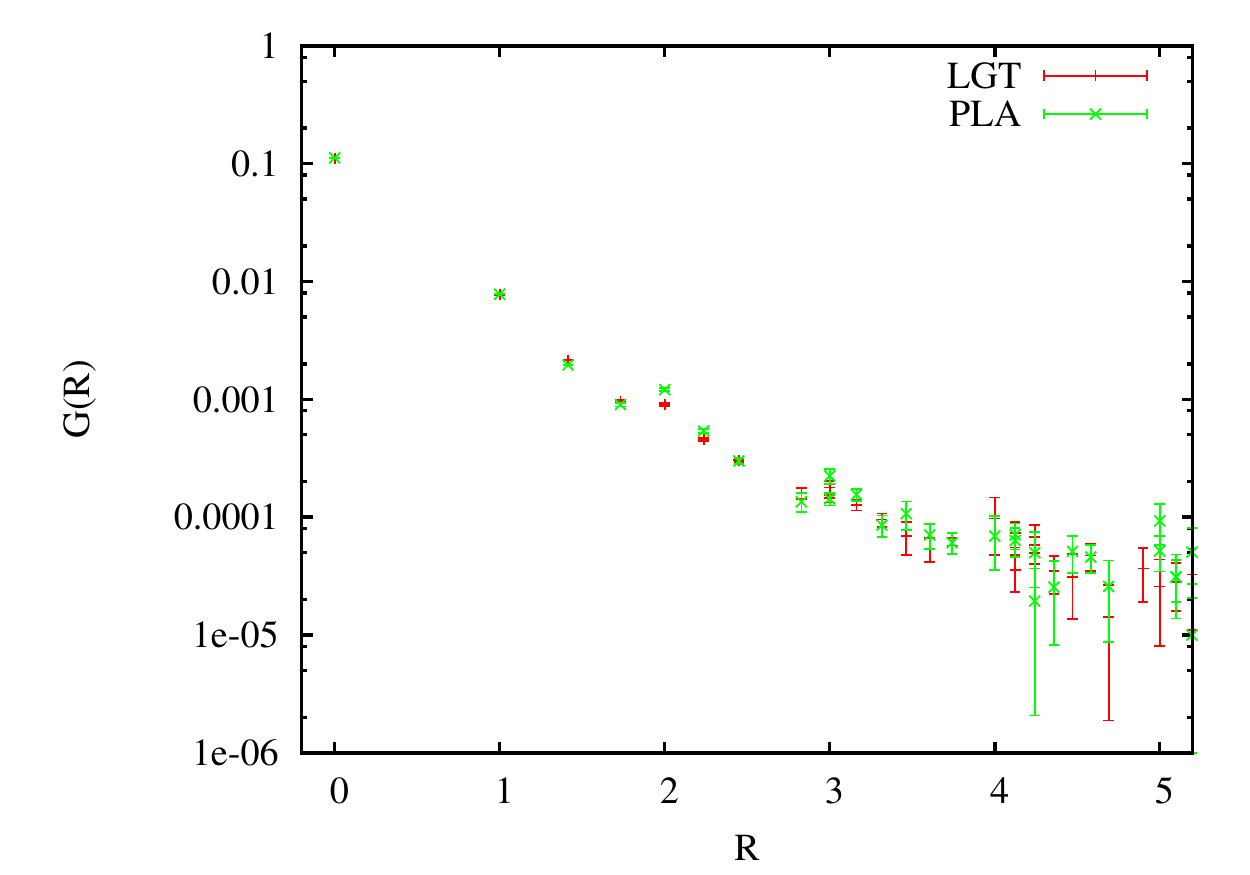}
\includegraphics[width=.49\linewidth]{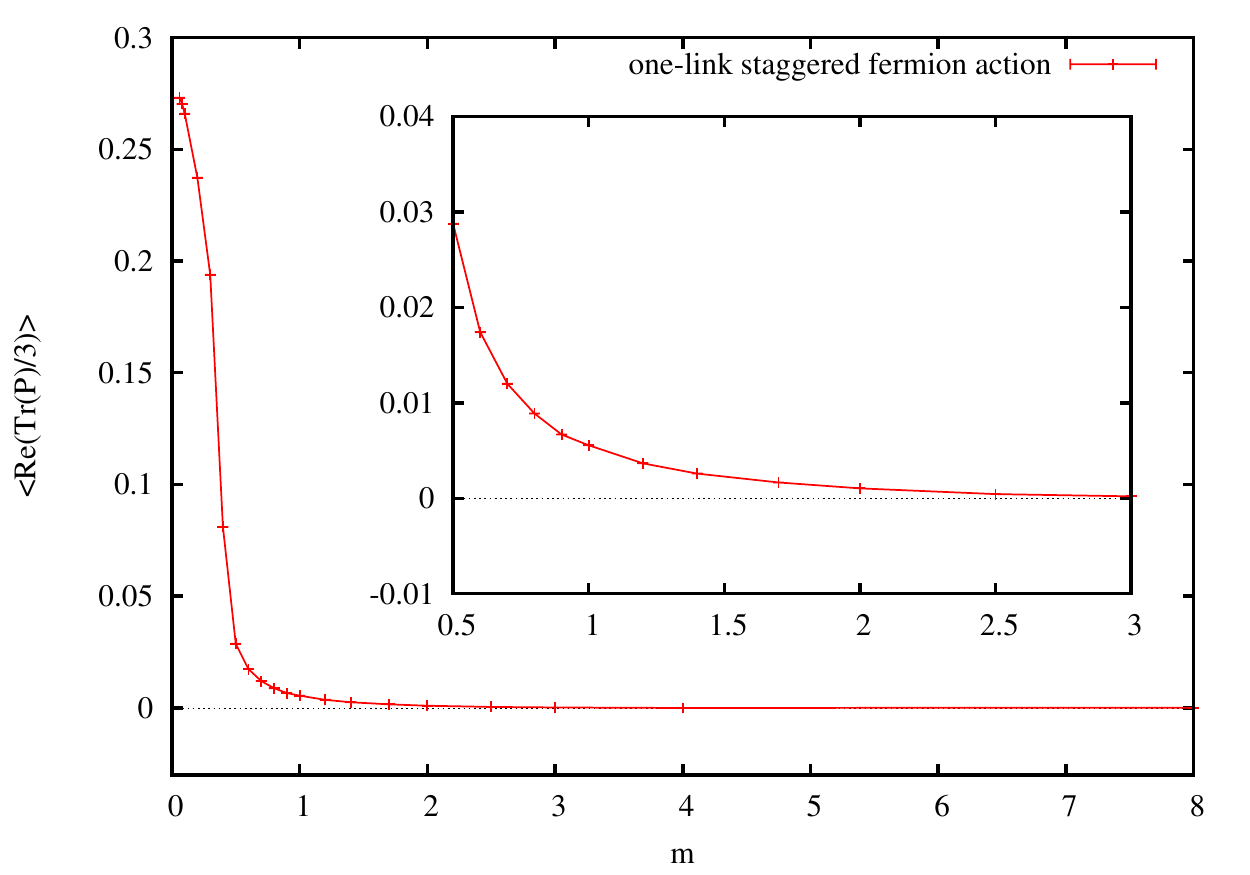}
\caption{(a) The Polyakov line correlators computed from numerical simulation of the 
effective PLA $S_P$, and from simulation of the underlying lattice SU(3) gauge
theory at $\beta=5.4$ with staggered fermions of mass $ma=1$ on $16^3\times4$ lattices. (b) The VEV of the Polyakov lines in the gauge theory, vs.\ mass $m$ in lattice units.}
\label{fig:placor}
\end{figure}

Now that we have obtained the kernel $K(\vx-\vy)$ and $h$ we can simulate the
effective PLA (\ref{eq:SP}) at $\mu=0$ by standard lattice Monte Carlo methods. We calculate the 
correlator
\beq
             G(R) = {1\over N^2} \langle \tr[P_\vx] \tr[P_\vy^\dg] \rangle ~~,~~ R=|\vx-\vy|
\label{GR}
\eeq
and compare it with the corresponding Polyakov line correlator computed in the
underlying SU(3) lattice gauge theory at $\beta=5.4$ with staggered fermions of
mass $ma=1$ on $16^3\times4$ lattices. The comparison (including off-axis separations)
is shown in Fig.\ \ref{fig:placor}a). Allowing for the fact that the data is a
little noisy beyond $R=4$, this seems like good agreement.
Fig.\ \ref{fig:placor}b) shows the expectation value (VEV) of the Polyakov lines in
the lattice gauge theory at $\beta=5.4$, showing a phase transition below
$ma=1$. The PLA simulation reproduces the correct Polyakov correlators (and
VEV) in the higher mass region above the phase transition, however, going to smaller quark
masses we find discrepancies, indicating that our ansatz for $S_P$ is inadequate beyond the
transition/crossover.

\section{Preliminary Results at Finite Density}

\begin{figure}
\centering
\includegraphics[width=.49\linewidth]{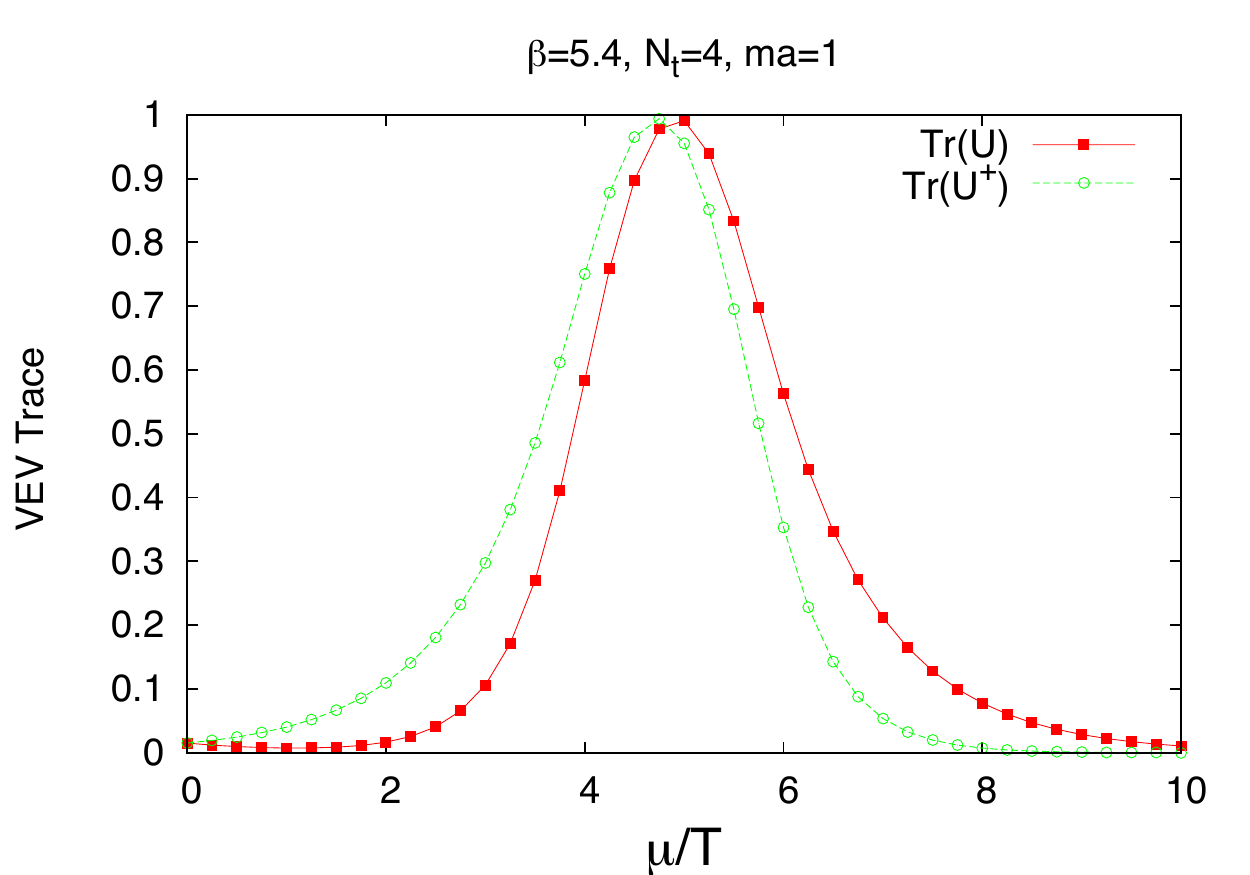}
\includegraphics[width=.49\linewidth]{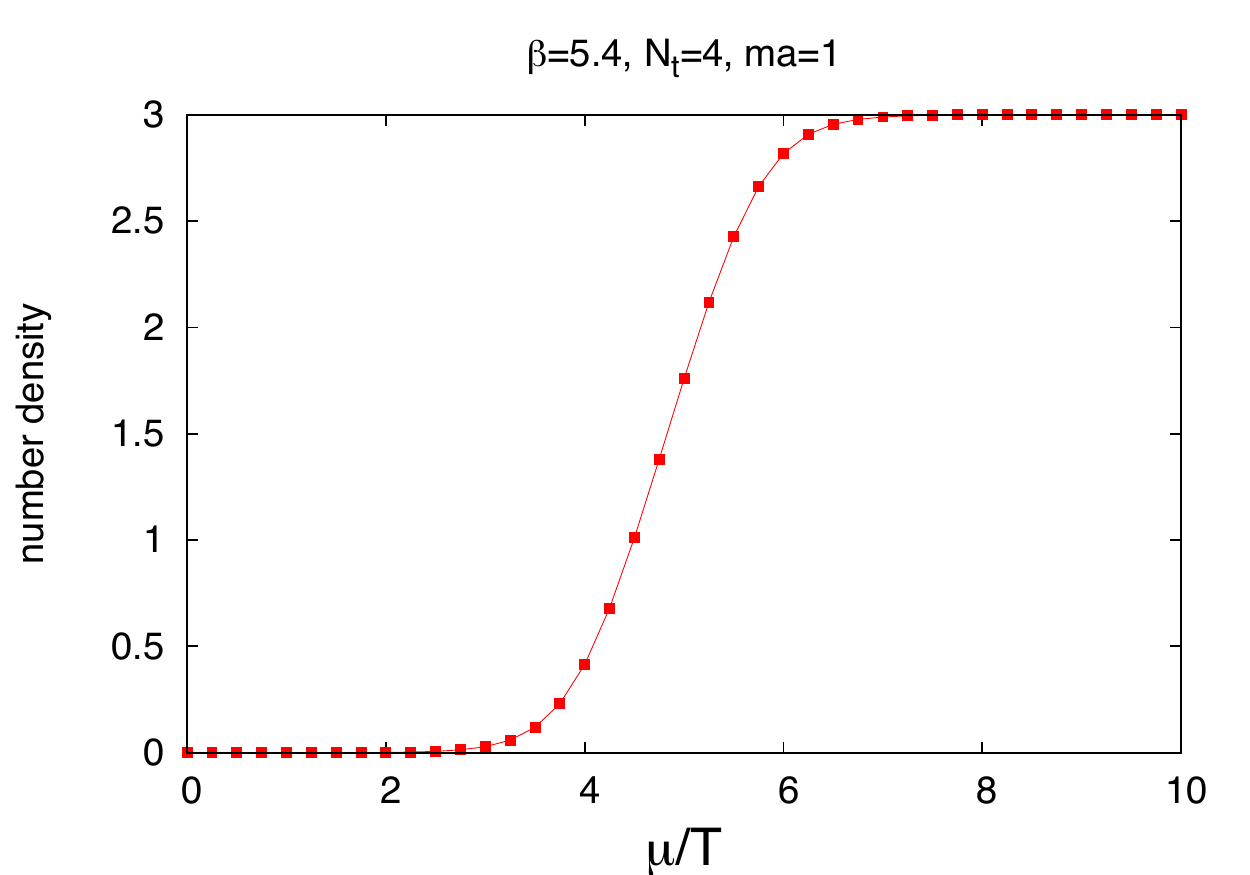}
\caption{Mean field solution of the effective Polyakov line action $S_P$
	corresponding to a gauge theory at $\beta=5.4$ with staggered quarks of mass
	$ma=1$, $16^3 \times 4$, at finite values of the chemical potential.  (a)
the expectation value of Polyakov lines $\langle \tr U \rangle$ and $\langle \tr
U^\dg \rangle$ vs.\ $\mu$; (b) particle number density vs.\ $\mu$.}
	\label{fig:res}
\end{figure}

At finite $\mu$ there is still a sign problem in the effective action (3.4), and
this problem we address via the mean field approach developed in~\cite{Greensite:2012xv}. The  treatment of $SU(3)$ spin models at finite $\mu$ is a minor variation of standard mean field theory at zero chemical potential.  For details of the approach, and
comparison to complex Langevin see~\cite{Greensite:2012xv}. The results of the
mean field calculation at $\beta=5.4$ and $N_t=4$ for staggered quarks of mass
$ma=1$, using the extracted parameter $h$, are shown in Fig.~\ref{fig:res}. 
The VEVs of Polyakov lines $\langle \tr U \rangle$ and $\langle \tr U^\dg
\rangle$ in Fig.~\ref{fig:res}b) reproduce the correct value at $\mu=0$ and go
to $0$ for large $\mu/T$, while the number density in Fig.~\ref{fig:res}b)
saturates at $n=3$ particles/lattice site, as required by the Pauli exclusion
principle for staggered quarks with three color. There is no phase transition
for $ma=1$, as in the heavy quark case.

\section{Conclusions \& Outlook}

We have presented the relative weights method for extracting the effective Polyakov
line action from SU(3) lattice gauge theory at $\beta=5.4$ with dynamical
staggered fermions of mass $ma=1$ on $16^3\times4$ lattices. We find good agreement between Polyakov line correlators computed in the effective action and in the underlying gauge theory at zero chemical potential. Mean field methods have been employed to determine the expectation value of observables in the effective action as a function of chemical potential.
    
The next step is to proceed to lighter quarks. In this case we may have to fit
our data to a more complicated action. If the results at $\mu=0$ pass our usual tests, then
we will attempt to locate phase transitions in the QCD phase
diagram, which is the ultimate goal.

\bibliographystyle{../utphys}
\bibliography{../literatur}

\end{document}